\documentclass[10pt,conference]{IEEEtran}
\pdfoutput=1

\usepackage{algorithmic}
\usepackage{amsmath,amssymb,amsfonts,amsthm}
\usepackage{cite}
\usepackage[colorlinks,urlcolor=black,linkcolor=black,citecolor=black,linktocpage=true]{hyperref}
\usepackage{cleveref}
\usepackage{comment}
\usepackage{lipsum}
\usepackage{graphicx}
\usepackage{siunitx} 
\usepackage{subfigure}
\usepackage{textcomp}
\usepackage{todonotes}
\usepackage{xcolor}
\usepackage{xspace}
\usepackage{svg}
\usepackage{multirow}
\usepackage[normalem]{ulem}
\usepackage{verbatim}
\usepackage{url}
\usepackage{makecell}
\usepackage{listings}

\lstnewenvironment{terminal}[1][]
{
    \lstset{
        basicstyle=\fontsize{6}{7}\selectfont\ttfamily,
        breaklines=true,
        breakatwhitespace=true,
        columns=flexible,
        keepspaces=true,
        showspaces=false,
        showstringspaces=false,
        frame=none,
        xleftmargin=0pt,
        xrightmargin=0pt,
        aboveskip=2pt,
        belowskip=2pt,
        basewidth=0.45em,
        #1
    }
}{}

\useunder{\uline}{\ul}{}

\newtheoremstyle{compact}
  {0.5ex}{0.5ex}{\slshape}{}{\bfseries}{.}{0.5em}{}\theoremstyle{compact}
\newtheorem{researchquestion}{RQ}
\crefname{researchquestion}{RQ}{RQs}

\newcommand{\SWH}{Software Heritage\xspace}
\newcommand{\ourtool}{\textsc{\small GitHistorian}\xspace}

 \newcommand{\FullDataset}{2024-08-23\xspace} 

\newcommand{\SWHOriginShort}{\num{310}\,M\xspace}

\newcommand{\OriginsTwoVisitsOrMoreShort}{\num{111}\,M\xspace}

\newcommand{\OriginsOneVisitOrLessShort}{\num{199}\,M\xspace}

\newcommand{\PercentOriginsAltered}{\num{1.10}\%\xspace}\newcommand{\AlteredCommitsShort}{\num{12.5}\,B\xspace}

\newcommand{\GitRootCauseCommitsShort}{\num{8.7}\,M\xspace}
\newcommand{\GitOriginsRootCauseCommitsShort}{\num{1.22}\,M\xspace}

\newcommand{\SecretsRemovedAllShort}{\num{13}\,M\xspace}
\newcommand{\SecretsOriginsAllShort}{\num{75}\,k\xspace}
\newcommand{\SecretsRemovedIdRsa}{\num{108}\xspace}
\newcommand{\SecretsRemovedIdRsaOrigins}{\num{23}\xspace}

\newcommand{\SecretsRemovedAllKeyShort}{\num{6.9}\,M\xspace}
\newcommand{\SecretsRemovedAllSecretShort}{\num{805}\,k\xspace}

\newcommand{\LicenseAll}{\num{796972}\xspace}
\newcommand{\LicenseRemoved}{\num{719196}\xspace}
\newcommand{\LicenseModified}{\num{77776}\xspace}
\newcommand{\LicenseModifiedBothScanned}{\num{65688}\xspace}
\newcommand{\LicenseUpdate}{\num{79}\%\xspace}
\newcommand{\LicenseFullChange}{\num{14}\%\xspace}
\newcommand{\LicensePartialChange}{\num{5}\%\xspace}
\newcommand{\LicenseSeveralFamilies}{\num{8024}\xspace}
\newcommand{\LicenseOrigins}{\num{32169}\xspace}
\newcommand{\LicenseOriginsThousandPlusStars}{\num{76}\xspace}

\newcommand{\FileAlterationDepth}{\num{10}\xspace}

\newcommand{\AlteredCommits}{\num{12 542 848 352}\xspace}

\newcommand{\GitRootCauseCommits}{\num{8 720 085}\xspace}
\newcommand{\GitOriginsRootCauseCommits}{\num{1 218 547}\xspace}

\newcommand{\OriginsUnkownStars}{\num{283 203}\xspace}
\newcommand{\OriginsZeroOneStar}{\num{676 252}\xspace}
\newcommand{\METARootCauseCommits}{\num{1 160 725}\xspace}
\newcommand{\DIRRootCauseCommits}{\num{6 693 247}\xspace}

\newcommand{\PercentMETARootCauseCommits}{\num{13.3}\%\xspace}
\newcommand{\PercentDIRRootCauseCommits}{\num{76.8}\%\xspace}

\newcommand{\CategFileModified}{\num{5 469 080}\xspace}
\newcommand{\CategFileRemoved}{\num{1 287 304}\xspace}
\newcommand{\CategContentSplit}{\num{763 369}\xspace}
\newcommand{\CategDifferentBranchName}{\num{866 113}\xspace}
\newcommand{\CategOther}{\num{55 638}\xspace}
\newcommand{\CategAuthor}{\num{667 424}\xspace}
\newcommand{\CategCommitter}{\num{649 937}\xspace}
\newcommand{\CategCommitterDate}{\num{864 607}\xspace}
\newcommand{\CategDate}{\num{769 287}\xspace}
\newcommand{\CategMessage}{\num{691 060}\xspace}

\newcommand{\OriginsThousandPlusStars}{\num{17 726}\xspace}

\newcommand{\OriginsTwoNineStars}{\num{115 431}\xspace}
\newcommand{\OriginsTenNinetyNineStars}{\num{82 519}\xspace}
\newcommand{\OriginsHundredNineHundredNinetyNineStars}{\num{43 416}\xspace}

\newcommand{\PercentOriginsZeroOneStar}{\num{72.3}\%\xspace}
\newcommand{\PercentOriginsTwoNineStars}{\num{12.3}\%\xspace}
\newcommand{\PercentTenNinetyNineStars}{\num{8.8}\%\xspace}
\newcommand{\PercentHundredNineHundredNinetyNineStars}{\num{4.6}\%\xspace}
\newcommand{\PercentThousandPlusStars}{\num{1.9}\%\xspace}

\IEEEoverridecommandlockouts
\begin{document}

\title{Altered Histories in Version Control System Repositories: Evidence from the Trenches\thanks{This work was supported by France Agence Nationale de la Recherche (ANR), program France 2030, reference ANR-22-PTCC-0001. This work was made possible by Software Heritage, the universal source code archive:\linebreak \url{https://www.softwareheritage.org}.}}

\author{\IEEEauthorblockN{Solal Rapaport\IEEEauthorrefmark{1},
    Laurent Pautet\IEEEauthorrefmark{2},
    Samuel Tardieu\IEEEauthorrefmark{3},
    Stefano Zacchiroli\IEEEauthorrefmark{4}}
  \IEEEauthorblockA{LTCI, Télécom Paris, Institut Polytechnique de Paris,
    Palaiseau, France\\
    Email:
    \IEEEauthorrefmark{1}solal.rapaport@telecom-paris.fr,
    \IEEEauthorrefmark{2}laurent.pautet@telecom-paris.fr,
    \IEEEauthorrefmark{3}samuel.tardieu@telecom-paris.fr,\\
    \IEEEauthorrefmark{4}stefano.zacchiroli@telecom-paris.fr}
}

\maketitle

\begin{abstract}
  Version Control Systems (VCS) like Git allow developers to locally rewrite recorded history, e.g., to reorder and suppress commits or specific data in them.
  These alterations have legitimate use cases, but become problematic when performed on public branches that have downstream users: they break push/pull workflows, challenge the integrity and reproducibility of repositories, and create opportunities for supply chain attackers to sneak into them nefarious changes.
  
  We conduct the first large-scale investigation of Git history alterations in public code repositories.
  We analyze \OriginsTwoVisitsOrMoreShort (millions) repositories archived by Software Heritage, which preserves VCS histories even across alterations.
  We find history alterations in \GitOriginsRootCauseCommitsShort repositories, for a total of \GitRootCauseCommitsShort rewritten histories.
  We categorize changes by where they happen (which repositories, which branches) and what is changed in them (files or commit metadata).

  Conducting two targeted case studies we show that altered histories recurrently change licenses retroactively, or are used to remove ``secrets'' (e.g., private keys) committed by mistake.
  As these behaviors correspond to bad practices---in terms of project governance or security management, respectively---that software recipients might want to avoid, we introduce \ourtool, an automated tool, that developers can use to spot and describe history alterations in public Git repositories.
\end{abstract}

\section{Introduction}
\label{sec:intro}

Distributed version control systems (DVCS), like Git~\cite{ProGit2014, DBLP:books/daglib/0029843}, are among the most popular tools used by software developers~\cite{online:stack-survey-2022}.
In a typical Git workflow, each developer works on a local project repository that contains the full development history and integrate code changes there, by adding new commits.
Then, to share changes with collaborators, developers push new commits to public repositories and pull from there new commits done by teammates into their local repository.
These back and forth commit exchanges continue as long as the project is active.
The consolidated view of the project is available at any time from a canonical public repository, which is used by users and other downstream recipients to obtain the software source code.

For the most part Git repositories are append-only: developers add \emph{new} commits to their repositories.
But it is technically possible to \emph{alter the previously recorded history} of a Git repository~\cite[Chapter 7.6 Git Tools --- Rewriting History]{ProGit2014} and modify old commits.
Two common reasons for altering Git histories are: (1) ``amending'' a recent commit to change its content or metadata, and (2) ``rebasing'' a portion of VCS history to merge commits, split them, or avoid merges by ``linearizing'' development branches.
Any change to a previously recorded commit results in one or more commits changing their public identifiers
(cf.~\Cref{sec:background} for details).

History rewrites are useful to polish local and/or work-in-progress VCS branches before sharing it with others, and are even the default behavior in some development workflows~\cite{DBLP:journals/infsof/RiosEE22}.
For example, pull request (PR) branches are often rewritten in order to keep only a single commit in them that can be merged into the target branch when the PR is ready.

History alterations are however problematic when performed on public branches used by downstream recipients, for various reasons.
First, they break the pull/push workflow~\cite{DBLP:journals/cscw/NguyenI18, DBLP:conf/issre/JiCYM20}, which relies on the stability of commit identifiers as checkpoints for history exchanges.
Second, by breaking commit identities, alterations create opportunities for supply chain attacks on repositories~\cite{ohm2020backstabber, ladisa2023supplychain}: together with a legitimate history rewrite, attackers can sneak in malicious changes, which will be difficult to audit due to the confusion caused by commit identity modifications.

\subsection{Contributions}

Little is currently known about the \textbf{amount, characteristics, and impact of VCS history alterations in public code}.
This is partly because it is an inherently difficult phenomenon to study empirically at scale: after rewrite, the VCS history \emph{prior} to the alteration no longer exists and cannot be compared with what remains post-alteration.
This work contributes to bridge this gap, by conducting the first large-scale quantitative and qualitative analysis of history alterations in public VCS repositories.
Specifically, we address the research questions detailed below.
\begin{researchquestion}\label{rq:how-much}
  How often are the histories of publicly available VCS repositories altered in an observable way?
\end{researchquestion}
To solve the methodological problem of having access to VCS histories before and after alterations, we mine \SWH (SWH)~\cite{swhipres2017, swhcacm2018}, the largest public archive of software source code, which maintains full copies of public code repositories, even across history rewrites.

We analyze \OriginsTwoVisitsOrMoreShort (millions) public Git repositories archived from major development forges like GitHub as well as many GitLab public instances.
We identify \AlteredCommitsShort (billions) altered histories, detected by the observable that at least one commit reachable from a previously archived repository state is missing from a later one.
The phenomenon is quantitatively non-negligible: it impacts \GitOriginsRootCauseCommitsShort repositories.

\begin{researchquestion}\label{rq:where-changes}
  What (a) branches and (b) repositories are impacted the most by history alterations?
\end{researchquestion}
We breakdown the quantitative findings from \Cref{rq:how-much} by branches, to understand how the phenomenon relates to development workflows; and by repository popularity, to understand if underused and potentially lower-quality repositories are more impacted or not.

We then turn our attention to qualitative aspects:
\begin{researchquestion}\label{rq:what-changed}
  When the history of a VCS repository is altered, which commit \emph{parts} undergo modifications?
\end{researchquestion}
We categorize what is altered using a novel taxonomy that captures whether files (i.e., versioned files) or metadata (e.g., timestamps, commit messages, etc.) are changed.
We find that \PercentMETARootCauseCommits history alterations concern metadata-only changes, in most cases impacting multiple metadata at once, whereas \PercentDIRRootCauseCommits alterations include changes to files and/or directories.

\begin{researchquestion}\label{rq:change-patterns}
  Are there recurrent patterns of file alterations among observed VCS history rewrites?
\end{researchquestion}
To evaluate concrete risks for software developers, we look into the raw results obtained from \Cref{rq:what-changed} for recurring patterns of files that are frequently part of history alterations.

We analyze two specific scenarios: (a) removal of ``secrets'', like private keys, and (b) license changes occurring in appropriately-named files (e.g., \texttt{LICENSE}).
Instances of (a) denote subpar security diligence on the part of the developers; (b) instances are also problematic because, while license changes can legitimately happen during the lifetime of a project, they should not happen \emph{retroactively} by altering VCS history, as that might results in users who lack old copies of the repository losing rights.
To help developers spotting these and other problematic history alteration patterns, we explore one final question:
\begin{researchquestion}\label{rq:tool}
  Can we design and implement an automated tool to audit and detect history alterations in public repositories?
\end{researchquestion}
We demonstrate how our tool \ourtool performs both efficient and accurate analysis on very large-scale archival datasets.
It enables developers to check if repositories of their interest underwent history alterations, providing detailed information about (and optional filtering on) what was changed, when the alterations occurred, and which commits were affected.
This enables project maintainers to quickly detect suspicious past alterations and investigate further.
\ourtool can also be integrated into CI/CD pipelines to provide automated alerts upon detection of history alterations.

\subsection{Data availability statement}
A full replication package containing the data and code for the experiments presented in this paper, as well as the \ourtool tool, is available from Zenodo and Software Heritage~\cite{replication-package}.

\section{Background}
\label{sec:background}

\paragraph{Git data model}

The data model of Git, like most other DVCS~\cite{brindescu2014dvcs}, is a Merkle~\cite{Merkle} direct acyclic graph (DAG), where nodes are used to represent commits and other types of source code artifacts.
Each DAG node is identified by a cryptographic checksum computed recursively on its content and metadata which, for commit nodes, include the identifiers of previous commits.
The full VCS history of a project hence forms a connected graph whose integrity can be verified efficiently by only considering the identifiers of outer ``root'' nodes: if they match a previous known state, then all previous commits reachable from them have not been altered since.

\paragraph{History alteration with Git}

\begin{figure}
  \centering
  \includegraphics[width=\linewidth]{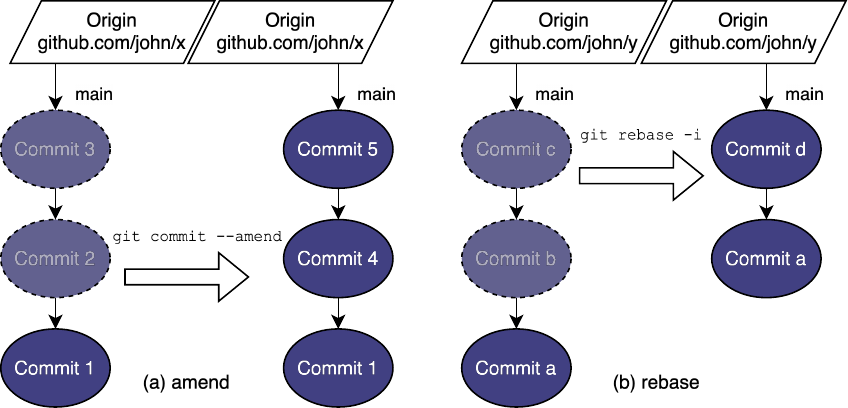}
  \caption{Common scenarios leading to history alterations in Git: (a) commit amendment, (b) interactive ``rebase''. Grayed out commits are no longer reachable after alteration. Arrows point from newer to older commits.}
    \label{fig:git-history-rewrite}
\end{figure} 

\Cref{fig:git-history-rewrite} shows two common use cases of history rewrites in Git: (a) amending and (b) rebasing.
The ``amend'' functionality allows developers to retroactively modify parts of commits that were already recorded in the version history:
\emph{commit metadata} (e.g., timestamp, message, author name or email, etc.) and/or \emph{commit content} (e.g., adding or removing files).
In \Cref{fig:git-history-rewrite}(a), the user modifies commit \texttt{2}, producing the new commit \texttt{4}.
After history alteration, due to how Merkle identifiers work, commit \texttt{2} disappears from the graph while commit \texttt{4} is added to it.
In the example, since commit \texttt{3} is based on commit \texttt{2}, commit \texttt{3} is \emph{also} modified and becomes commit \texttt{5}.
History alteration can therefore induce a ``\emph{snowball effect}'', where all the commits that transitively reference modified ones get new identities as well.
If commit \texttt{2} had one million commits (transitively) referencing it instead of one, they would \emph{all} disappear from the repository and be replaced by new ones.

\Cref{fig:git-history-rewrite}(b) shows the other common use case of Git history rewrites: ``rebasing''.
In this example, commits \texttt{c} and \texttt{b} are merged into a single commit: they both disappear and are replaced by new commit \texttt{d}.
In other use cases, rebase can also split commits or attach them to different previous commits.

\paragraph{Terminology}

A \emph{repository snapshot} (or simply \emph{state}) is the observable state of a repository at a given point in time.
Intuitively, it corresponds to the set of all commits in it.

We say that a repository $R$ underwent a \emph{history alteration} if there exist two subsequent repository snapshots $S_1, S_2$ of $R$ such that $\exists c_i, c_i \in S_1 \land c_i \not \in S_2$, i.e., at least one commit is present in the first snapshot but missing from the second.
Multiple history alterations can be observed for the same repository if multiple snapshot pairs $S_i, S_{i+1}$ denote history alterations.

An \emph{altered commit} is a commit impacted by a history alteration, as detected by the fact that is missing from a subsequent repository snapshot, but present in a previous one.
Due to the snowball effect, potentially many commits will be altered as a consequence of a single history alteration, but a minimum of one commit must be (otherwise the alteration will not be detected).

For each history alteration, we identify one or more \emph{root cause commits} as the oldest commits, in topological order, among all altered commits.
Root cause commits are commits that have been \emph{directly} altered: their identity changes are not a mere consequence of the modification of previous commits.

\paragraph{\SWH} 

\begin{figure}
  \centering
\includegraphics[width=\linewidth]{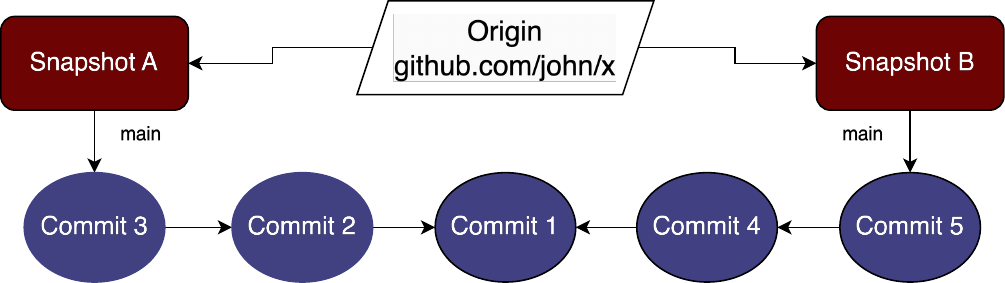}
  \caption{Repository state before and after \texttt{\small git commit --amend} of \Cref{fig:git-history-rewrite}(a).
    Both histories are preserved by \SWH and share common commits.
  }
  \label{fig:swh}
\end{figure} 

(SWH)~\cite{swhipres2017, swhcacm2018} archives the development history of more than 370 million projects hosted on major development forges like GitHub and GitLab.\footnote{\url{https://archive.softwareheritage.org}, accessed 2025-05-23.}
Each archived project is identified by its \emph{origin}: the repository URL.
SWH crawlers periodically visit each origin, taking complete snapshots of the repository state and storing collected artifacts in a global Merkle DAG.
As shown in \Cref{fig:swh}, if commits disappear, e.g., due to history alterations, from a repository between two archival visit, they will still be reachable from the \emph{previous} snapshot of the same origin.
We rely on this feature as the basis for our empirical experiments, whose methodology we describe next.

\section{Methodology}
\label{sec:methodology}

To answer the stated research questions, we followed an empirical methodology comprised of 4 phases:
\begin{enumerate}
\item we collect the largest existing corpus of periodically crawled public repositories for analysis (addressing \Cref{rq:how-much});
\item second, we devise an experimental protocol to detect history alterations (addressing \Cref{rq:how-much} and \Cref{rq:where-changes});
\item we create a taxonomy to categorize history alterations and understand the reasons behind them (\Cref{rq:what-changed});
\item based on the observed recurrent patterns of file name alterations, we illustrate the benefits of our approach and dig further into two case studies: history alterations to remove ``secrets'' committed by mistake, and alterations impacting license files (\Cref{rq:change-patterns}).
\end{enumerate}

\subsection{Data collection and repository selection}

In order to analyze a large amount of public software repositories, we start from the \SWH (SWH) graph dataset~\cite{DBLP:conf/msr/PietriSZ19}.
We retrieved the most recent dataset version available at the time of our experiments: timestamp
\FullDataset.\footnote{\url{https://docs.softwareheritage.org/devel/swh-dataset/graph/dataset.html}}
It contains the VCS history of more than \SWHOriginShort (millions) repositories and packages (``origins'' in SWH terminology).
We will detect history alterations by comparing successive snapshots of the same origin, as detailed below in  \Cref{subsec:framework}.
To avoid under-estimating the ratio of repositories that witness history alterations, we exclude those visited by SWH crawlers less than 2 times (\OriginsOneVisitOrLessShort).
We also exclude origins that are not Git repositories (0.019\,M), due to the popularity of Git and in order to better relate findings to practices.
This leaves us with \OriginsTwoVisitsOrMoreShort origins where we can potentially find evidence of history alterations. 

To answer \Cref{rq:where-changes}(b), we will group repository by popularity and use rely on GitHub ``stars'' as popularity indicator~\cite{DBLP:journals/jss/BorgesV18}.
Although the number of stars can be artificially inflated by malicious actors~\cite{DBLP:journals/corr/abs-2412-13459}, it is still a relevant metric in our case: a large number of stars increases repository visibility, making it more worthy of scrutiny, including to detect history alterations.
We rely on GitHub star counts made available by SWH, that also crawls platform-specific metadata like these.
Due to the uneven popularity of development platforms, in the following we restrict analyses that depend on popularity to GitHub repositories; other analyses are conducted on all repositories archived by SWH, independently of the platform.

\subsection{General framework}
\label{subsec:framework}

\begin{figure}
  \centering
  \includegraphics[width=\linewidth]{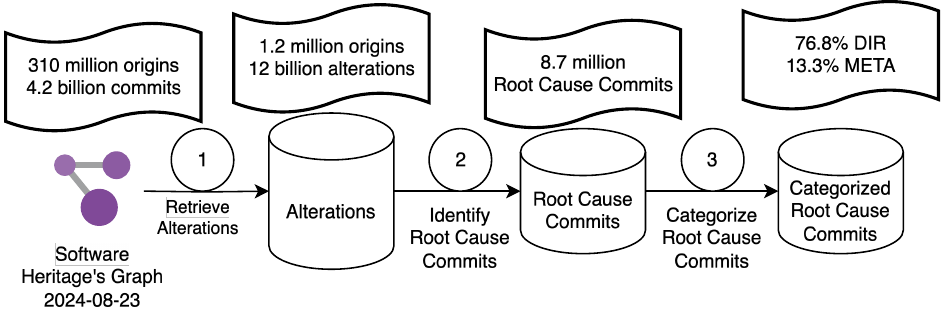}
  \caption{Methodology for detecting and categorizing history alterations.}
  \label{fig:methodology}
\end{figure}

\Cref{fig:methodology} shows the experimental approach we devised to detect and categorize history alterations.
For each SWH origin with at least two visits, and for each Git branch in them, we compare all pairs of successive snapshots to identify commits from the earlier snapshot which are missing from the later one. In this set of commits, we identify the \emph{root cause commits} according to the definition given in \Cref{sec:background}.
We implement this step in practice with custom code that mine locally the compressed graph representation of the SWH archive~\cite{saner-2020-swh-graph}, via its Rust API.
This detection framework enables us to address both \Cref{rq:how-much} (by counting alterations) and \Cref{rq:where-changes} (by analyzing branch-level patterns).

\begin{table}
  \caption{Branch names unification}
  \label{tab:branch-names-unification}
  \centering
\begin{tabular}{|l|l|}
    \hline
    \textbf{Branch names or patterns} & \textbf{Unified name} \\
    \hline
    \ttfamily main, master & \emph{main} \\\hline
    \ttfamily dev, devel, develop, development & \emph{development} \\\hline
    \texttt{pull}/\emph{number}/head & \emph{pull request} \\\hline
    \texttt{renovate}/\emph{anything} & \emph{renovate} \\
    \hline
  \end{tabular}
\end{table}

In Git, different branch naming schemes are used to represent the same concept.
For example, branches \texttt{main} and \texttt{master} are both used to designate the main stable branch of a repository, while branches hosting GitHub pull requests start with \texttt{pull/}.
A popular bot used to maintain project dependencies up-to-date uses the \texttt{renovate/} prefix, and development branches often start with \texttt{dev}.
We unify together branches used for the same purpose, to provide aggregate statistics in the following (addressing \Cref{rq:where-changes} (a)).
Details about how we unified branches together are shown in \Cref{tab:branch-names-unification}.

\subsection{History alteration detection and categorization}
\label{subsec:categ}

\begin{table}
    \centering
    \caption{Root cause commits by category}
    \label{tab:res-all}
    \begin{tabular}{|ll|r|r|}
    \hline
    \multicolumn{2}{|l|}{\textbf{Category}}                                             & \textbf{\#Commits} & \multicolumn{1}{c|}{\textbf{Total}}         \\ \hline
    \multicolumn{1}{|c|}{\multirow{6}{*}{\textbf{META}}} & \textbf{Author}              & \CategAuthor           & \multirow{6}{*}{\METARootCauseCommits} \\ \cline{2-3}
    \multicolumn{1}{|c|}{}                               & \textbf{Message}             & \CategMessage            &                          \\ \cline{2-3}
    \multicolumn{1}{|c|}{}                               & \textbf{Date}                & \CategDate            &                          \\ \cline{2-3}
    \multicolumn{1}{|c|}{}                               & \textbf{Committer}           & \CategCommitter           &                          \\ \cline{2-3}
    \multicolumn{1}{|c|}{}                               & \textbf{Committer Date}       & \CategCommitterDate           &                          \\ \cline{2-3}
    \multicolumn{1}{|c|}{}                               & \textbf{Other}               & \CategOther             &                          \\ \hline
    \multicolumn{1}{|l|}{\multirow{3}{*}{\textbf{DIR}}}  & \textbf{File Modified}     & \CategFileModified           & \multirow{3}{*}{\DIRRootCauseCommits} \\ \cline{2-3}
    \multicolumn{1}{|l|}{}                               & \textbf{File Removed}      & \CategFileRemoved           &                          \\ \cline{2-3}
    \multicolumn{1}{|l|}{}                               & \textbf{Content Split}      & \CategContentSplit            &                          \\ \hline
    \multicolumn{2}{|l|}{\textbf{Different Branch Name}}                               & \CategDifferentBranchName           & \CategDifferentBranchName           \\ \hline
    \end{tabular}\end{table}

We categorize the root cause commits of history alterations using the custom taxonomy shown in \Cref{tab:res-all} (we found no preexisting taxonomy for this in the literature), which allows to capture alterations to both metadata and file (and/or directory) content.
This taxonomy directly addresses \Cref{rq:what-changed} by systematically characterizing what commit parts undergo modifications.
The taxonomy was developed via an iterative mixed method: sampling uncategorized history alterations, manually comparing their before/after states, capturing the individual data that differed (leading to introducing additional categories if needed), applying all categories to the entire dataset, and repeating until all data was categorized.
Starting from the simplest case: a (root cause) commit is categorized as \emph{Different Branch Name} if its content and metadata have \emph{not} been altered, but it is now found only in a different branch than the previous one.

A commit is categorized as \emph{META} it has been replaced by a new commit with strictly identical files and directories \emph{content}, but different metadata.
This categorization is further refined to indicate which metadata fields have been altered.
Multiple metadata fields can be modified in a single history alteration.

A commit is categorized as \emph{DIR} if the commit file/directory content was altered or deleted, in part or as a whole.
In this case, we do not analyze metadata changes, as there is no single reference commit to compare them with.
The DIR categorization is further refined to describe what happened to altered content:
\begin{itemize}
\item \emph{Content Split}: the content of each file that existed in the altered commit is still in the repository, but is now split into several other commits. This is exclusive of other subcategories in ``DIR''.
\item \emph{File Modified}: at least one file has been modified.
\item \emph{File Removed}: at least one file was removed, and was not found in other commits.
\end{itemize}

To distinguish among these we look in the newest snapshot at commits starting from the children of the parents of the root cause commit, and proceed for \FileAlterationDepth generations (commit tree depth).
If, after \FileAlterationDepth generations of descendants, or if there are no more descendants, one of the file paths cannot be found, the file is considered removed and the commit is marked as such.
We chose \FileAlterationDepth generations heuristically, as a trade-off between analysis time and diminishing likelihood that a file will reappear at the same path without being a distinct file.

\subsection{Case study design}
\label{sec:methodology-case-studies}

With the taxonomy developed thus far we are able to peek into recurring patterns of history alterations at various levels, including which files are modified retroactively in the DIR category.
Based on this, we looked more in-depth into two recurring patterns of VCS history alteration, addressing \Cref{rq:change-patterns}.

\paragraph{Secrets suppression}
The first case study focuses on secrets removed (category: File Removed) from a repository by altering its history.
With ``secret'' we mean potentially sensitive information that should never have been shared via a public repository.
We detect this based on popular file names used for storing private keys of passwords, like \texttt{id\_rsa} (without \texttt{.pub} extension), \texttt{secring.gpg}, etc.
The full list is available from the replication package~\cite{replication-package}; further examples in Section \ref{sec:casestudy-secrets}.

\paragraph{Retroactive license changes}
The second case study identifies retroactively modified license files on the main branch (after branch unification) of a repository.
It is considered bad practice to \emph{retroactively} alter the license of an open source project, as it makes impossible for novel users to obtain the software under its previous license, possibly to fork it.

To detect license changes, we look at altered files (category: File Modified) containing \texttt{LICENSE} (or \texttt{LICENCE}, ignoring case) in their name.
We then run ScanCode~\cite{scancode-home, DBLP:journals/computer/Ombredanne20}, a state-of-the-art license detection tool, on the files before and after alteration, considering only the highest-confidence result and only if it is above 90\%.
We obtain this way two \emph{sets} of licenses (before/after) that we further categorize as follows to evaluate the magnitude of the change:
\begin{itemize}

\item \emph{License Update}: the set of unversioned licenses (e.g., GPL, Apache) remain the same, whereas that of versioned licenses changes (e.g., GPL-2 $\to$ GPL-3).
  Note that, even in this lower-impact case, there is no valid reason to make the change \emph{retroactively} by rewriting Git history.

\item \emph{Partial Change}: some (unversioned) licenses have changed, e.g., GPL has been added, MIT removed.

\item \emph{Full Change}: \emph{all} (unversioned) licenses have changed.

\end{itemize}

\section{Results}
\label{sec:results}

\subsection*{\Cref{rq:how-much}: how often are VCS histories altered?}

Across our entire corpus, we observed \AlteredCommits alterations spread across all analyzed branches.
In terms of repository count, \GitOriginsRootCauseCommits contained at least one altered commit, representing approximately \PercentOriginsAltered of all examined repositories.
While the proportion is modest in relative terms, it is not negligible in absolute ones: users of more than 1\,M repositories can---and will in the future assuming future stability---use public Git repositories that underwent history alterations during their lifetime.

At the same time, the low percentage indicates that history rewriting remains an exception rather than standard practice, suggesting that common Git usage practices adhere to the expectation of immutable version control histories.

\subsection*{\Cref{rq:where-changes}a: what branches are altered the most?}

\begin{figure}
    \includegraphics[width=\columnwidth]{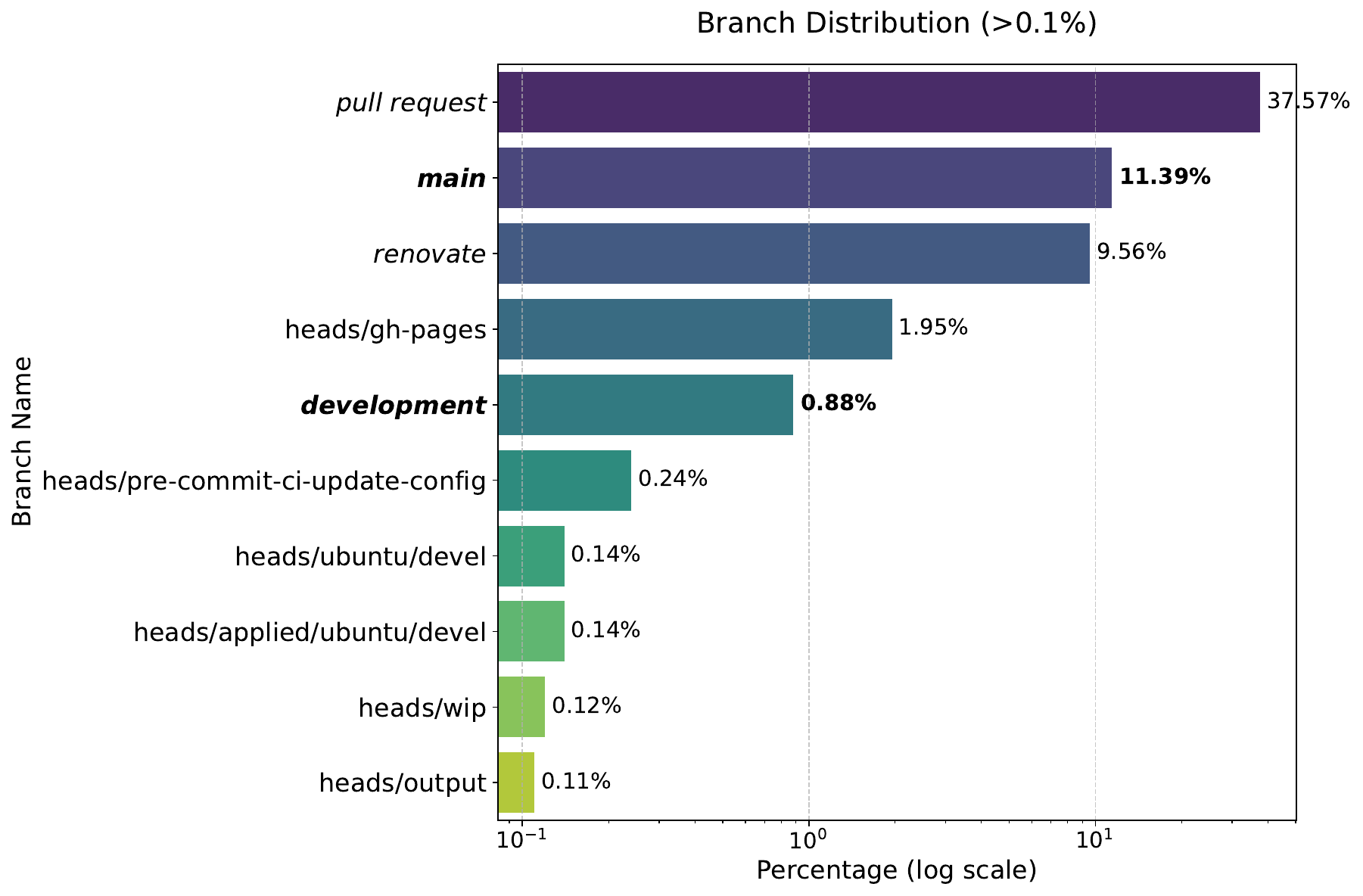}
    \centering
    \caption{Most impacted branches by Altered Histories ($> 0.1$\%)}
    \label{fig:distribBranch}
\end{figure}

We now look into where history alterations happen, starting within repositories at the granularity of specific branches.
For this analysis we consider branch names after the unification of \Cref{tab:branch-names-unification}.

\Cref{fig:distribBranch} presents the distribution of history alterations by branch, showing the top-10 branches.
The results reveal some patterns that align with popular development expectations, but also highlight more concerning cases.
Indeed, as one might expect, pull request branches represent the most frequent target of history alterations, with $37.57\%$ of all observed alterations.
This pattern reflects the common practices where contributors refine their pull requests through interactive rebasing, commit squashing, etc.~to produce a clean history before integration.

Other high-ranked branches give cause for concern.
\emph{Main} branches (after unification, hence spanning both \texttt{main} and \texttt{master}) are in general expected to be long-lived and stable, but still experienced history alterations in $11.39\%$ of the cases we observed, ranking as the 2nd-most impacted category.
Additionally, development branches account for $0.88\%$ of observed alterations, ranking 5th overall.
Depending on the project and workflow, development branches might be expected to be stable or not.

The 3rd-most impacted category at $9.56\%$, \emph{renovate} branches, represents automated dependency update workflows.
These alterations stem from bot behavior that rewrites commits when dependency proposals evolve prior to merging, making history modification an inherent characteristic of automated maintenance processes.

Even though, around $50\%$ of history alterations follow a known practice, these findings raise significant implications for repository integrity and developer trust.
History alterations on main branches compromise the fundamental assumptions of version control systems, potentially breaking downstream dependencies, invalidating signed commits, and disrupting reproducible builds~\cite{DBLP:conf/acmrep/CourtesSZT24}.
The $12\%$ alteration rate for main branches suggests that a substantial subset of repositories may pose reliability risks for users who depend on stable commit references.

\subsection*{\Cref{rq:where-changes}b: what repositories are altered the most?}

\begin{figure}
    \includegraphics[width=\columnwidth]{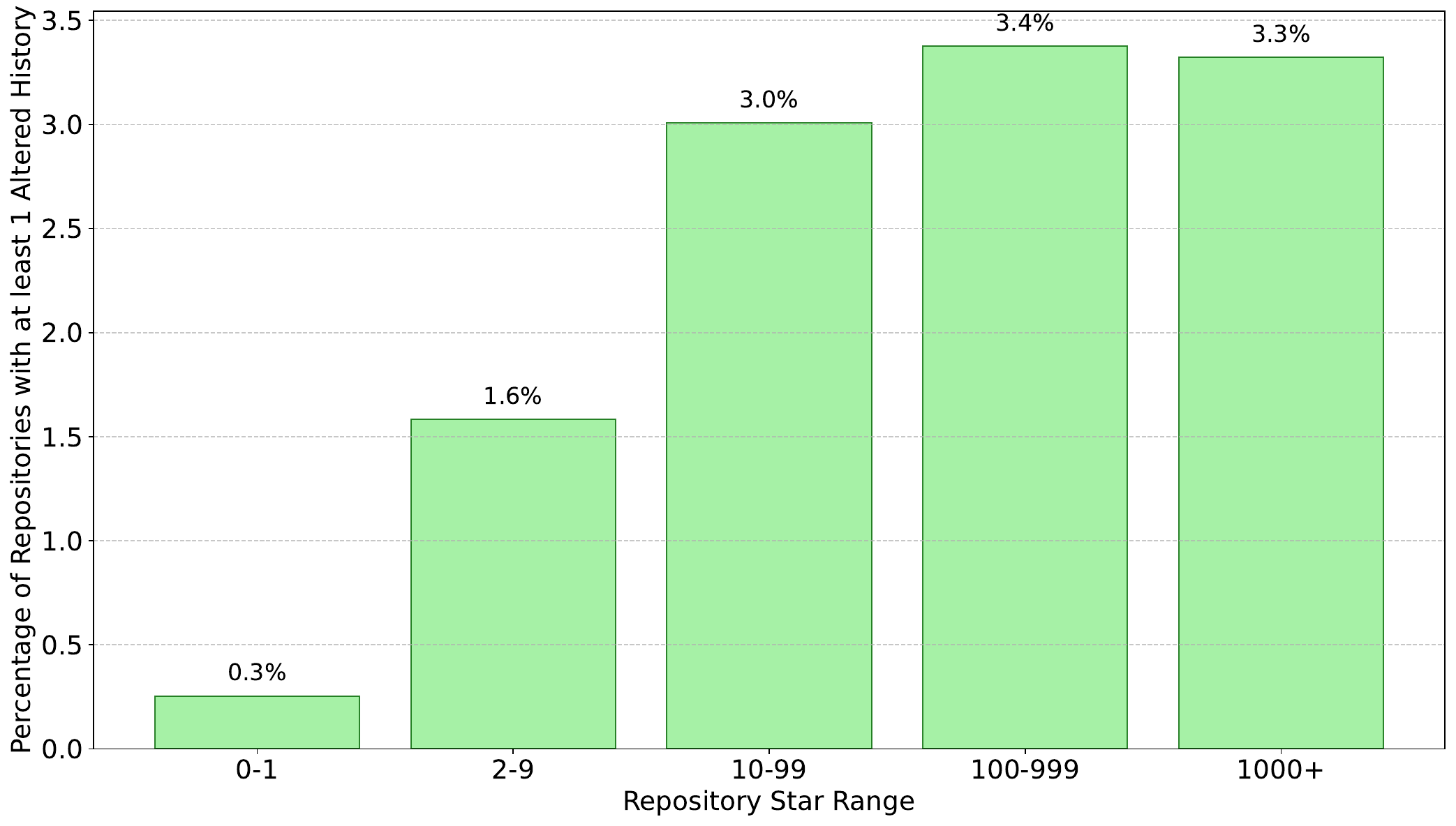}
    \centering
    \caption{Proportion of Repositories with Altered Histories per Star Range}
    \label{fig:resStar}
\end{figure}

Not all repositories are born equal: repositories created by student just because their teacher said so do not have as many downstream users as \texttt{torvalds/Linux.git}.
One might then wonder if repository popularity correlates with the amount of history rewrites, under the hypothesis that maturity leads to a more limited use of disruptive VCS practices.
To explore this we join the set of repositories that experienced at least one alteration with their amount of GitHub stars.
From \GitOriginsRootCauseCommits altered repositories, we remove \OriginsUnkownStars origins with unknown amount of star.
We observe the following star distribution:
\begin{itemize}
  \item 0 to 1 star: \OriginsZeroOneStar origins (\PercentOriginsZeroOneStar)
  \item 2 to 9 stars: \OriginsTwoNineStars origins (\PercentOriginsTwoNineStars)
  \item 10 to 99 stars: \OriginsTenNinetyNineStars origins (\PercentTenNinetyNineStars)
  \item 100 to 999 stars: \OriginsHundredNineHundredNinetyNineStars origins (\PercentHundredNineHundredNinetyNineStars)
  \item 1000 and plus stars: \OriginsThousandPlusStars origins (\PercentThousandPlusStars)
\end{itemize}
\Cref{fig:resStar} shows the percentage of origins with detected history alteration for each bucket.
The analysis reveals a pronounced relationship between repository popularity and the likelihood of history alterations.
Repositories with minimal popularity (0 to 1 star) exhibit the lowest alteration rate at $0.3\%$.
This rate increases to $1.6\%$ for repositories with 2 to 9 stars, $3.0\%$ for those with 10 to 99 stars, and then stabilizes at approximately $3.4\%$ for repositories with 100 to 999, $3.3\%$ for those with 1000 and plus stars.

It is hence \emph{not} the case that less popular repositories experience more rewrites than more popular ones.
Rather, as repositories gain visibility, the likelihood of history modifications increases substantially.
Even very mature and popular repositories (100+ stars) experience a non-negligible amount of history alterations, which is potentially concerning in terms of repository auditability.

We also looked into which branches are altered the most and compared results with repository popularity.
In very popular repositories (100+ stars) history alterations happen for the most part in pull request branches ($56\%$) but remains non negligible in \emph{main} branches ($1.8\%$).
This finding has significant implications for dependency management, reproducible builds, and the general reliability assumptions that developers make when incorporating external repositories into their workflows, irrespectively of their popularity.

\subsection*{\Cref{rq:what-changed}: which parts of commits are altered?}

\begin{figure}
    \includegraphics[width=\columnwidth]{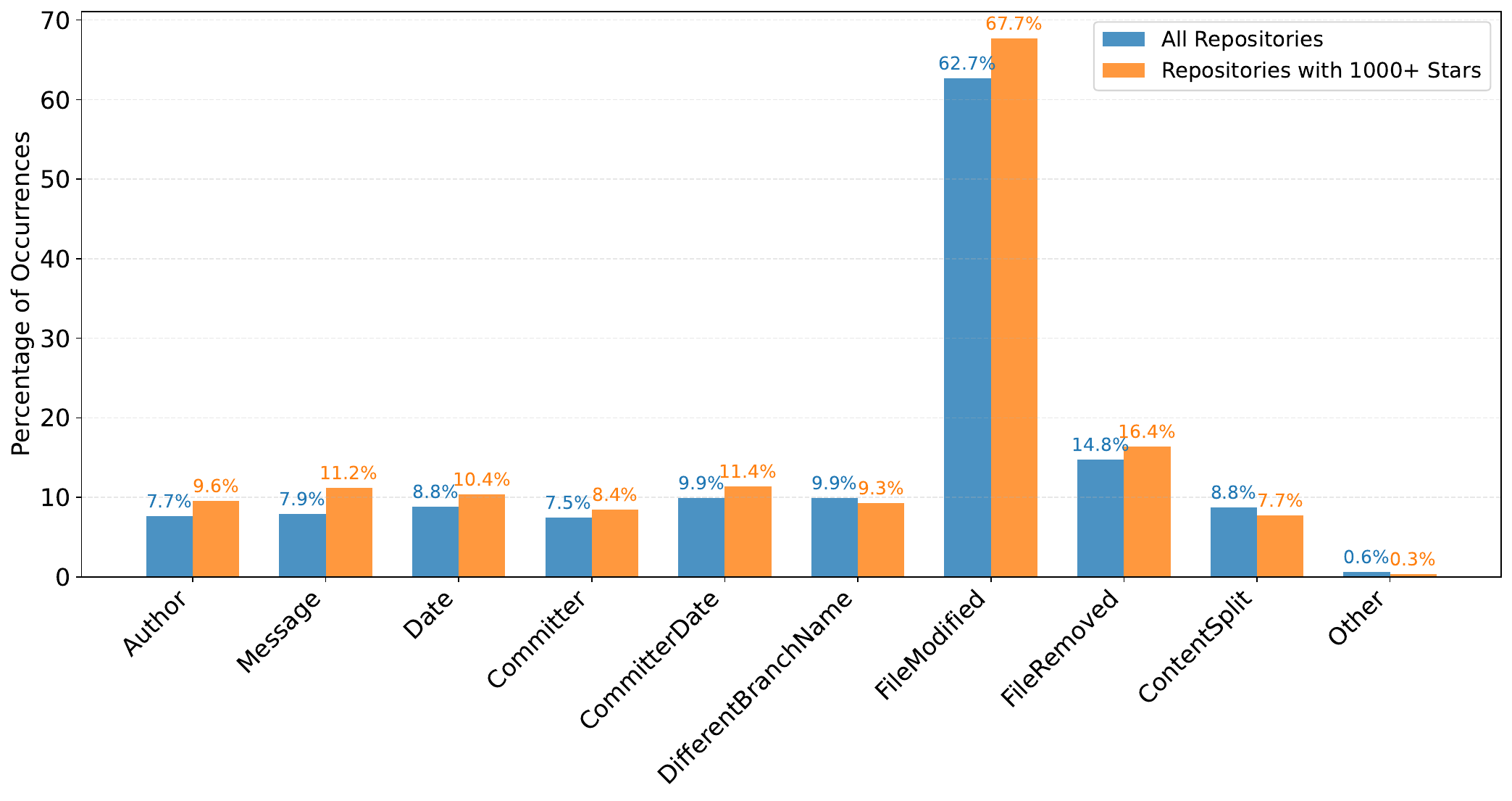}
    \centering
    \caption{Distribution of Categories: All vs. Popular Repositories}
    \label{fig:distribCateg}
\end{figure}

We apply the taxonomy of \Cref{tab:res-all} looking first for metadata-only changes (category: META), then for file/directory changes (category: DIR).
The results for each category are displayed in \Cref{fig:distribCateg}, which shows the proportion of altered histories in each category---META categories on the left, DIR categories on the right---for both the entire corpus and for the subset of most popular GitHub repositories (1000+ stars).

\subsubsection{Metadata attributes changes (META)} 

Among the \GitRootCauseCommits identified root cause commits, \METARootCauseCommits instances (\PercentMETARootCauseCommits) exhibited modifications only to commit metadata, leaving underlying files and directories unaltered.

Digging into specific metadata categories shows that committer date represents the most frequently modified attribute across both the complete dataset and the subset of repositories with 1000 or more stars.
We do not observe significant variations in the modification patterns of the two sets: metadata modification practices remain uniform regardless of popularity.

We also conducted a more in-depth analysis of the co-occurrences of metadata changes, i.e., META cases that involve changes to 2 or 3 metadata fields at the same time.
The most frequent alteration for 2 fields at once is the change of author date, which always coexist with a change to the committer date---not doing so would indeed be surprising.

The most frequent triplet of metadata fields changed together in the global dataset that occurs 85\% of the time is when the commit message changes; in most of those cases author and committer dates change as well.
This result corresponds to the expected behavior when modifying a commit message.

\subsubsection{Directory content changes (DIR)}

We categorized \DIRRootCauseCommits altered commits based on changes to the content of the files or directories.
The dominance of the category \emph{File Modified} is significant in both the global dataset and the subset of popular repositories that display a proportion higher than $60\%$. 
In both datasets the occurrence of files being modified in a history alteration is more than 4 times higher than the occurrence of files being removed.
Developers who alter a file in the history are 4 times more likely to modify rather than deleting, renaming or moving it around.

The final subcategory corresponds to commits moved between branches: \emph{Different Branch Name}, with \CategDifferentBranchName commits.
The most plausible explanation for this are merges of so-called ``feature branches'', which are later removed from the repository; this behavior is still problematic for direct users of feature branches, but are understandable if those branches are explicitly documented as volatile.

\subsection*{\Cref{rq:change-patterns}: are there recurrent file patterns in alterations?}

\begin{table}
\centering
\caption{Top 20 most altered file names}
\label{tab:top-files}
\resizebox{\columnwidth}{!}{\begin{tabular}{|l|l|r||l|l|r|}
\hline
\textbf{Rank} & \textbf{File Name} & \textbf{Count} & \textbf{Rank} & \textbf{File Name} & \textbf{Count} \\ \hline
1            & \path{default.nix}        & \num{13401935}       & 11           & \path{CHANGELOG.md}       & \num{5001971}        \\ \hline
2            & \path{package.json}       & \num{12594717}       & 12           & \path{__init__.py}    & \num{4859517}        \\ \hline
3            & \path{Makefile}           & \num{10615195}       & 13        & \path{metadata.xml}       & \num{4477947}        \\ \hline
4            & \path{README.md}          & \num{9969599}        & 14           & \path{CMakeLists.txt}     & \num{4380963}        \\ \hline
5            & \path{index.js}           & \num{9129894}        & 15           & \path{package.py}         & \num{4336900}        \\ \hline
6            & \path{Manifest}           & \num{8891876}        & 16        & \path{APKBUILD}           & \num{3375064}        \\ \hline
7            & \path{pom.xml}            & \num{7080355}        & 17           & \path{template}           & \num{3122049}        \\ \hline
8            & \path{index.html}         & \num{6297643}        & 18           & \path{BUILD}              & \num{3018131}        \\ \hline
9            & \path{meta.yaml}          & \num{5425074}        & 19           & \path{index.ts}           & \num{2764520}        \\ \hline
10           & \path{index.md}           & \num{5372929}        & 20           & \path{LICENSE}            & \num{2762147}        \\ \hline
\end{tabular}}
\end{table}

As we have seen while answering at \Cref{rq:what-changed}, file alterations are the most common forms of VCS history alteration.
This begs the question of \emph{which} files are being modified or removed.
To answer \Cref{rq:change-patterns} we extract the file name part of all file paths involved in history alterations of category DIR.
\Cref{tab:top-files} shows the top-20 of most commonly altered file names.

The most frequently altered files are predominantly configuration and build-related artifacts.
\path{default.nix} (\#1), \path{package.json} (\#2), \path{Makefile} (\#3), \path{pom.xml} (\#7), \path{CMakeLists.txt} (\#14), and \path{APKBUILD} (\#16) represent different build systems and package managers.
This suggests that dependency updates, version bumps, and build configuration changes are primary motivations for altering the history.

The high frequency of files like \path{default.nix}, \path{package.json}, and \path{meta.yaml} matches the workflows of automated dependency management tools (like \texttt{\small Renovate}, \texttt{\small Dependabot}~\cite{DBLP:journals/tse/HeHZZ23}, or \texttt{\small Nix update bots}) that systematically modify these files and subsequently rewrite commit histories when adjustments are needed.

Files like \path{README.md} (\#4), \path{CHANGELOG.md} (\#11), and \path{index.md} (\#10) represent project documentation, indicating that documentation improvements are another common reason for history rewrites.
The presence of entry points across different ecosystems (\path{index.js} \#5, \path{index.html} \#8, \path{index.ts} \#19, \path{__init__.py} \#12) suggests that main application files are frequently altered, reflecting refactoring or structural changes that seem to require frequent history cleanups.

Lastly, the appearance of \path{LICENSE} (\#20) and \path{metadata.xml} (\#13) indicates that legal compliance and project metadata corrections constitute a notable category of history alterations, reflecting retroactive license changes or potential metadata correction efforts.

This pattern analysis reveals that history alterations predominantly target infrastructure, configuration, and metadata files rather than core application logic, suggesting that maintenance activities and automated tooling are primary contributors to the observed history alterations.

\paragraph{Case study: removing ``secrets'' from history}
\label{sec:casestudy-secrets}

Across all repositories, we identified \SecretsRemovedAllShort history alterations categorized as \emph{File Removed}, involving files whose names commonly denote private information that should generally not be distributed publicly, like private keys, private certificates, and password.
These alterations occur across \SecretsOriginsAllShort (thousands) different repositories in our dataset.

Generic naming patterns demonstrate the highest frequency of occurrence, with files containing ``key'' and ``secret'' accounting for \SecretsRemovedAllKeyShort and \SecretsRemovedAllSecretShort altered files, respectively.
Manual examination of a representative random sample confirmed that removed files contained sensible private information, in files such as \path{secret.yaml} (containing authentication credentials) and \path{samlKey.jks} (storing keys for the deployment of production environments).

SSH private keys were less popular, probably due to the fact that platforms like GitHub can nowadays block Git pushes containing them, but are still not absent from our dataset: we identified \SecretsRemovedIdRsa removed files containing RSA private keys, spanning \SecretsRemovedIdRsaOrigins origins.

These findings demonstrate that despite developers' attempts to remediate secret exposure through history alteration, such sensitive information remains recoverable through history digging in archives.
When developers inadvertently publish private information, removing it from \emph{their} repository history represents only partial remediation.
Complete security restoration requires regenerating or rotating all exposed tokens, passwords, and cryptographic keys.
But if they do so, there is arguably little point in rewriting the repository history afterwards.

It is important to remind here that via our methodology we can only detect alterations occurring between captured repository snapshots.
Consequently, our results correspond to either cases where developers were unfortunate enough to have snapshots captured immediately following secret commits, or instances where significant time elapsed between exposure and remediation attempts.

\paragraph{Case study: retroactive license changes}
\label{sec:casestudy-licenses}

In our second case study we focus on retroactive license modifications, where developers alter VCS history changing the content of license-denoting filenames (e.g., \texttt{LICENSE}).
These changes are quite frequent in our experiments, as evidenced by the presence of \path{LICENSE} among the top-20 most frequently altered file names in \Cref{tab:top-files}.

We identified \LicenseAll altered license files across the main branches of our dataset, across \LicenseOrigins different repositories, with \LicenseOriginsThousandPlusStars of them having 1000+ stars on GitHub.
Among these, \LicenseRemoved instances involved file renaming, relocation, or deletion, while \LicenseModified cases represented content modifications.

Using ScanCode as described in \Cref{sec:methodology-case-studies}, we successfully identified the set of before/after-alteration licenses for \LicenseModifiedBothScanned history alterations involving license files.
Our analysis reveals that \LicenseUpdate of alterations are relatively minor license updates, implicating at most changes in license versions.
Further manual analysis indicates that copyright modifications (e.g., in the name of the author) are a common occurrence in this category.
\LicenseFullChange of alterations represent full license changes, encompassing both transitions from more permissive (e.g., MIT) to more restrictive (e.g., GPL) licenses and vice-versa.
Only \LicensePartialChange of alterations constitute partial changes, which we attribute to the limited number of repositories (\LicenseSeveralFamilies alterations) detected with multiple license families, thereby reducing the representativeness of partial change scenarios.

These findings are concerning for downstream users of open source products, because while \emph{legally} most open source software licenses are irrevocable and meant to grant specific rights to users forever, benefiting of those rights require practical access to a version of the software released under a given license.
Retroactively changing version history can constitute an attempt to make old software versions, under an old license, disappear from public circulation, inhibiting previously available rights.
This is even more concerning for industrial users, who tend to appreciate, and expect, legal stability in the open source software they depend on.

\section{Detecting and avoiding altered histories}

VCS history alterations go unnoticed by everyone, except for downstream users who happen to have retrieved and kept a repository version before the alteration, and also \texttt{git pull} after the alteration, with branch tracking in place for the affected branches.
Few users satisfy these conditions.
For example, it is not uncommon for CI/CD pipelines that depend on external software to perform a fresh \texttt{git clone} of repositories of interest, making it impossible to detect history alterations.
Nonetheless, some history rewrite patterns might be causes of concerns for downstream users of affected repositories: they might hint at bad maintenance practices or, worse, be evidence of active tampering by malicious actors.
To help both downstream users and upstream maintainers worried about history alterations, we answer \Cref{rq:tool} practically, by designing, implementing, and showcasing \ourtool: a practical tool capable to detect, describe, and avoid (if desired) repositories that witnessed VCS history alteration.

\subsection{Design}

The design of \ourtool satisfies three core requirements: scalability for analyzing thousands of repositories, accuracy in detecting various types of history alterations, and usability for both interactive analysis and automated monitoring.
The system architecture consists of four main components: (1) a PostgreSQL database storing preprocessed history alteration data derived from \SWH snapshots,\footnote{The local database is required because SWH currently does not provide a remote API answering the queries needed to detect history alterations. This requirement can be lightened in the future by either SWH operators adding this service, or by some third parties providing it, consuming the datasets that SWH publish periodically, as we did one-off for our experiments.} (2) a command-line interface (CLI) for interactive repository analysis, (3) a caching mechanism to optimize repeated queries, and (4) Git hook integration for continuous monitoring of repository changes.

The detection mechanism operates by querying the database for known alterations.
The tool supports branch-specific analysis, allowing users to focus on main development branches or examine all branches comprehensively.

For automated monitoring, the tool employs Git hooks that trigger checks after repository updates (\texttt{\small post-merge}) and branch switches (\texttt{\small post-checkout}).
This design ensures that developers are immediately notified when working with repositories that have experienced history alteration, without requiring manual intervention.

\subsection{Implementation and CLI}

\ourtool is implemented in Rust, using the \texttt{\small sqlx} crate for type-safe database interactions and \texttt{\small tokio} for asynchronous operations, enabling efficient handling of database queries even with large datasets like ours.
The implementation is available as part of the replication package of this paper~\cite{replication-package}.

The main functionalities can be accessed via the main CLI commands:

\textbf{Database Management:} The \texttt{\small load} command handles the initial dataset ingestion into the local database, with parallel processing support.
This allows the efficient loading of datasets containing billions of alteration records.

\textbf{Repository Analysis:} The \texttt{\small check} command requests auditing a specific repository to detect history alterations.
It queries the local database as needed, and formats results for human consumption.
It is possible to analyze main branches only, development branches only, or all branches.

\textbf{Automated Monitoring:} The \texttt{\small attach} command installs Git hooks that call the \texttt{\small check-cached} command.
This cached variant implements smart result caching based on repository URL, target branch, and dataset version, avoiding redundant database queries whenever possible.

\textbf{Caching Strategy:} Results are cached by repository URLs, with cache validity determined by dataset version comparison.
This approach ensures that results remain accurate when new alteration data becomes available while providing immediate responses for repeated queries if desired, or no response to avoid redundancy.

Hook scripts include contextual information about the triggering event (pull, merge, or checkout) and provide clear user feedback about the analysis results.

\subsection{Example}

Consider a developer working on a project who wants to ensure their repository dependencies did not undergo history alteration.
After installing and loading data into \ourtool, one can verify the absence of alterations in any repository of interest as follows:
\begin{lstlisting}
$ git-historian check https://github.com/example/project --branch main --verbose
Connected to the database!
Found 2 altered history records for 'https://github.com/example/project'

Altered History Records:
Record #1:
  Branch Name: refs/heads/master
  Altered Commit: swh:1:rev:a1b2c3d4e5f6789...
  Snapshot Destination: swh:1:snp:abcd1234...
  Sub Category: FileModified

Record #2:
  Branch Name: refs/heads/dev
  Altered Commit: swh:1:rev:f6e5d4c3b2a1098...
  Snapshot Destination: swh:1:snp:efgh5678...
  Sub Category: CommitterDate

File Modifications:
Record #1:
  Branch Name: refs/heads/master
  Altered Commit: swh:1:rev:a1b2c3d4e5f6789...
  File Path: src/security/auth.py
  Status: Modified

Results saved to: ~/.local/state/git-historian/
  altered_history_example_project_20241201_143022.txt
\end{lstlisting}
The repository underwent two types of history alteration: a file modification and a metadata alteration to change the commit date.
The verbose output shows that a security-related file was modified in one of the altered commits, information that could be crucial for security auditing.

For automatic monitoring in the future upon clone or pull actions, the developer can attach the tool to their local repository (or CI stateful repositories) like this:
\begin{lstlisting}
$ cd /path/to/local/repo
$ git-historian attach . --branch main --verbose
  git-historian successfully attached to repository
  Repository: .
  Remote URL: https://github.com/example/project
  Hooks installed:
  - post-merge (triggered after git pull)
  - post-checkout (triggered after git checkout)
\end{lstlisting}

\section{Discussion}
\label{sec:discussion}

\subsection{Implications for software development practices}

Our findings challenge the conventional wisdom about version control best practices in several ways. 
The paradoxical correlation between repository popularity and alteration frequency indicates that current development workflows may be fundamentally at odds with the immutability principle underlying version control systems.

While alterations on pull request and dependency management branches follow accepted practices, their normalization may be creating cultural acceptance that inadvertently extends to main branches.
This gradient of acceptability poses risks for repository integrity standards across the development lifecycle.

The substantial presence of main branch alterations represents a systemic challenge to fundamental software engineering assumptions.
Beyond the immediate technical impacts on CI/CD systems and semantic versioning, these practices undermine the cryptographic integrity guarantees that many security and compliance frameworks depend upon.
Organizations relying on commit signatures or audit trails may unknowingly operate with compromised assurance levels.

\subsection{Security and compliance implications}

Our secret removal analysis reveals deeper organizational security governance issues beyond immediate credential exposure.
The prevalence of post-hoc secret removal suggests systematic failures in preventive security measures, pre-commit hooks, developer training, and secure development workflows.
Organizations discovering secrets in their histories face a false choice between public exposure and historical integrity.

More concerning is the implicit assumption that history cleaning provides adequate remediation.
Our detection capabilities demonstrate that ``deleted'' secrets remain discoverable through historical analysis, yet many organizations may believe their incident response is complete after history rewriting.

The licensing case study exposes novel legal risks in open source governance.
Retroactive license changes through history alteration create ambiguous legal precedents where different project versions operate under potentially conflicting terms.
For organizations with complex supply chain dependencies, these alterations introduce uncertainty about intellectual property rights and compliance obligations that traditional license scanning cannot detect.

\subsection{Tool and ecosystem implications}

The automation-driven nature of many history alterations reveals an emerging disconnect between tool capabilities and user understanding.
Modern dependency management systems normalize history rewriting as an implementation detail, potentially creating user expectations that extend beyond appropriate use cases.
This normalization effect may contribute to the casual application of history alteration techniques in contexts where they are inappropriate.

Security tooling faces challenges in distinguishing between legitimate maintenance and potentially suspicious activity, requiring sophisticated context analysis beyond simple pattern matching.

The research ecosystem faces a methodological challenge as our findings demonstrate that significant portions of repository-based studies may analyze sanitized rather than authentic development data.
This has implications not only for historical accuracy but also for the reproducibility of software engineering research that relies on commit-based metrics and temporal analysis.

\subsection{Broader research and community implications}

These findings challenge several assumptions underlying current software engineering research and practice.
The prevalence of modifications suggests that a non-trivial portion of repository-based research may be analyzing sanitized or restructured data rather than authentic development histories.

For the broader open-source ecosystem, our findings raise questions about trust, reproducibility, and the long-term integrity of collaborative development platforms. The development of standards, best practices, and tooling to address these challenges represents an important area for future community investment.

\subsection{Supply chain attack obfuscation implications}

History alterations present a concerning attack vector for software supply chain compromise.
Attackers could inject malicious code while using history rewriting to erase evidence, making detection through conventional security auditing extremely difficult.
The normalization of history rewriting practices we documented creates perfect cover for malicious actors to make backdoors appear as legitimate maintenance activities.

The scale of alterations we observed---affecting millions of repositories---suggests this represents a significant blind spot in current supply chain security practices.
Traditional security measures focus on current repository states but may miss attacks exploiting the temporal dimension of version control history.

\section{Threats to validity}
\label{sec:threats}

\subsection{Construct validity}

History alteration is, by design, a destructive operation.
As such, the only way to observe it faithfully without information losses, would be on developer machines conducting a user study, dramatically reducing the scale of the experiment.
In this work we took the opposite approach of conducting a very large-scale experiment, relying on Software Heritage (SWH) to collect and preserve repository snapshots, which we then compare two-by-two to find traces of history alterations.

As such, we depend on the SWH archival frequency of public repositories: we can only observe history alterations that are visible between successive visits of a repository.
Malicious or accidental short-lived alterations might be missing from our analysis, if they have been fixed by reverting to an unaltered history version in between two archival visits of the same repository.
Also, we cannot distinguish \emph{multiple} history alterations that happen between two visits of the same repository: we will recognize them as a single alteration, merging together what changed in all of them.
To do better, one would need almost real-time archival of a large enough amount of public code repositories---something that, to the best of our knowledge, no dataset provides today---or, alternatively, an integrated insider study of a specific development platform (e.g., GitHub), which would introduce a selection bias.

Metadata changes are not detectable with our methodology when file changes \emph{also} occur, due to the lack of a commit to compare against.
As such, our analysis overlooks metadata changes that co-occurred with file ones.
We could nonetheless characterize them for more than 1\,M commits, but we make no claim about the rest.

When analyzing file changes we limit the search horizon to identify if and where previous content went, searching up to \FileAlterationDepth commits deep.
It is possible that content might be found with a larger horizon.
However, it would become debatable whether a file reappearing ``many'' commits away from a change is to be considered the same file or a new one.

\subsection{External validity}

We only considered Git repositories in this study and we do not claim further generality or applicability to other DVCS.
SWH archives code maintained using other DVCS as well, so it is theoretically possible to use the same dataset of this work to analyze others, possible for comparison purposes.
However, it would be difficult to disentangle the specificities of each DVCS, e.g., in terms of branch naming, for a proper comparison.
On the other hand, some DVCS allows for non-destructive history updates, which keeps track of previous history states.
Analyzing and comparing them with our findings would be interesting, even if by definition they would rule out some of the use cases considered in this work, like secret removal.

Finally, we observe that Git is a DVCS, but is also used as ``storage backend'' for other independent DVCS, like Jujutsu (\texttt{jj})~\cite{jujutsu-vcs-home}, where history alterations are used more liberally.
We did not attempt to detect and separate ``regular'' Git repositories from these cases.
This might impact our analysis, but we expect it to remain quantitatively marginal to this day.

\section{Related work}
\label{sec:related}

\subsection{History alterations: motivations and handling}

The Pro Git Book~\cite{ProGit2014} mentions history rewriting as a ``great thing'' and lists the actions one can perform \emph{before} sharing development history with others: modifying files, changing commit messages, reordering, squashing, splitting, or removing commits.
The ability to modify files is essential for removing erroneously committed sensitive information~\cite{DBLP:conf/secdev/BasakNRW22}.
Conversely, history alterations generate anomalies in the pull-based development model~\cite{DBLP:conf/wcre/Chouchen0KWTMM21, DBLP:conf/icse/Liu0MKYXGG24}, potentially disrupting collaborative workflows.

\paragraph*{Best practices and guidelines}
Developers use alterations to maintain ``clean'' histories, in order to provide a clearer overview of project evolution and facilitate decision-making during code reviews and project management.
Tools like, e.g., Githru~\cite{DBLP:journals/tvcg/KimKJKSKS21} encourage squashing to improve clarity.

Cortés Ríos et al.~\cite{DBLP:journals/infsof/RiosEE22} propose a taxonomy of workflows for collaborative software development, evaluating the pros and cons of approaches incorporating rebasing.
Other researchers have focused on recommending processes for pull-based development models, analyzing metrics such as commit quantity per pull request~\cite{DBLP:conf/ispw/AzeemPSS020} to guide merge decisions, and emphasizing that trustworthy software requires clean commit history~\cite{DBLP:conf/cms/PaulusMW13}.

Previous work looked into the challenges of resolving merge conflicts after rebases~\cite{DBLP:conf/issre/JiCYM20, DBLP:journals/cscw/NguyenI18, DBLP:conf/qrs/ShenYPZ23}, offering tools and recommendations for their resolution in collaborative workflows.

\paragraph*{Tools for detecting history alterations}
Limited tools exist to help recover missing elements from commit history, either through side-effects or inference techniques.
Abstract syntax tree (AST) based approaches~\cite{DBLP:conf/apsec/FujimotoHK21} attempt to detect file renames with high accuracy, focusing on content rather than metadata alterations.
Lavoie et al.~\cite{DBLP:conf/wcre/LavoieKMZ12} developed probabilistic methods to track file movements between versions, including addition, removal, and relocation, but they do not capture metadata alterations.

\subsection{Empirical analysis of history alterations}
There are almost no empirical studies on history alterations, probably due in part to the practical difficulty of conducting them (cf.~\Cref{sec:intro}).
The most relevant prior work was conducted by Germán et al.~\cite{DBLP:journals/ese/GermanAH16}, who subjected the official Linux kernel and its contributing repositories to continuous snapshots to better understand its development history.
Their findings suggest that the actual development history of projects using DVCS can be irretrievably lost.
While their study focused on alterations occurring across repositories, including rebasing and metadata changes, it did not examine alterations within individual repositories nor identify specific file-level changes, which we study at large-scale.

Previous work also identified the challenges of mining VCS histories for research purposes~\cite{DBLP:conf/msr/BirdRBHGD09, DBLP:journals/ese/KalliamvakouGBS16}, mentioning history rewrites as a significant concern.
Flint et al.~\cite{DBLP:journals/ese/FlintC022} studied timestamp anomalies in a small subset of SWH, suggesting that history alterations affect timestamp reliability, potentially compromising dataset integrity when mining software repositories.
They did not study other kinds of alterations.

\subsection{Research gap}
To our knowledge, no prior work has characterized, neither quantitatively nor qualitatively, the extent of history alterations in public code repositories, even less so at large-scale.
The present work fills this gap.
Furthermore, it provides a taxonomy to categorize history alterations, digs deeper in two recurrent patterns of alterations (secret suppression and retroactive license changes), and provides an automated tool to help developers who wish to be alerted in advance of the presence of history alterations in repositories of their interest.

\section{Conclusion} \label{sec:conclusion}

We conducted the first large-scale study of altered version control (VCS) histories in public repositories, performing both quantitative and qualitative analyses of history alterations in \OriginsTwoVisitsOrMoreShort (million) public repositories.
We found evidence of history alterations in \GitOriginsRootCauseCommitsShort (million) repositories.
The phenomenon is not negligible: due to it, downstream users of those repositories might have experienced disruption of their push/pull workflow, or might be currently using repositories that have been tampered with without their knowledge.
We have shown that history alterations are neither confined to unpopular repositories, nor to branches like pull request ones, where rewrites might be expected.

Applying a novel taxonomy for categorizing VCS history alterations, we discovered that most alterations involve changing the content of versioned files and directories.
The next occurrence in terms of incidence are metadata-only alterations, involving changes to author names, commit messages, timestamps, etc.
Conducting two targeted use cases we find that alterations are used to remove private information committed by mistake to repositories (e.g, private keys) and also to retroactively change the licensing of open source projects.
As users and developers might want to audit projects for history alterations---and given the difficulty of doing so: history alterations are designed to leave no traces---we develop \ourtool, a novel tool that leverage the \SWH archive to automatically spot alterations and alert users, possibly as part of CI/CD pipelines.

In future work, we intend to build upon the general framework of this paper to investigate the relationship between history rewrites and supply chain attacks.
That would require developing an approach to discern legitimate from malicious alterations and design automated detection systems for suspicious modification patterns.
We also plan to look more closely at the timing of rewrites, as it can provide important insights into developer awareness and security practices.
For instance, in the secret suppression scenario it would be interesting to distinguish early-stage removals, when the project was young and good security practices not mature yet, and late-stage removals which might be more worrisome.
More generally for all rewrites one could distinguish between immediate corrections---where rewrites happen within minutes or hours of the rewritten commit(s)---and long-term exposures, where an history of commits persists for days, months, or more before rewrite.

\section*{Acknowledgments}
The authors would like to thank Olivier Barais for his invaluable suggestions
on a preliminary version of this work.

\end{document}